\documentclass[times,sagev,times,review]{sagej}
\usepackage{moreverb,url}
\usepackage{mathrsfs}
\usepackage{amsfonts}
\usepackage{amsmath,bm}
\usepackage{amssymb}
\usepackage{graphicx}
\usepackage{epstopdf}
\usepackage{float}
\usepackage{subcaption}
\usepackage{verbatim}
\usepackage[
pdfstartview=FitH,
CJKbookmarks=true,
bookmarksnumbered=true,
bookmarksopen=true,
colorlinks, 
pdfborder=001,   
linkcolor=blue,
anchorcolor=blue,
citecolor=blue,
urlcolor=blue
]{hyperref} 
\newcommand\BibTeX{{\rmfamily B\kern-.05em \textsc{i\kern-.025em b}\kern-.08em
		T\kern-.1667em\lower.7ex\hbox{E}\kern-.125emX}}

\def\volumeyear{2016}

\begin{document}
	
\runninghead{Xiao, Meng, Dai, Zhang and Quan}

\title{A Lifting Wing Fixed on Multirotor UAVs for Long Flight Ranges}
\author{Kun Xiao\affilnum{1}, Yao Meng\affilnum{1}, Xunhua Dai\affilnum{1},
Haotian Zhang\affilnum{1} and Quan Quan\affilnum{1}}

\affiliation{\affilnum{1}School of Automation Science and Electrical Engineering, Beihang University, Beijing, China}

\corrauth{Quan Quan, Associate Professor, School of Automation Science and
	Electrical Engineering.Beihang University, Beijing 100191, China.}

\email{qq\_buaa@buaa.edu.cn}

\begin{abstract}
This paper presents a lifting-wing multirotor UAV that allows long-range flight. The UAV features a lifting wing in a special mounting angle that works together with rotors to supply lift when it flies forward, achieving a reduction in energy consumption and improvement of flight range compared to traditional multirotor UAVs. Its dynamic model is built according to the classical multirotor theory and the fixed-wing theory, as the aerodynamics of its multiple propellers and that of its lifting wing are almost decoupled. Its design takes into consideration aerodynamics, airframe configuration and the mounting angle. The performance of the UAV is verified by experiments, which show that the lifting wing saves 50.14\% of the power when the UAV flies at the cruise speed (15m/s).
\end{abstract}

\maketitle

\runninghead{Xiao, Meng, Dai, Zhang and Quan}

\section{Introduction}

\subsection{Lifting-wing multirotor UAV}

Nowadays, multirotor UAVs have been developing rapidly in consumer and
industrial markets owing to their advantages of vertical take-off and
landing, good maneuverability and stability, and simple configuration\cite{doi:10.1177/1756829317734835}. However, their operation range is poorer
than that of fixed-wing aircraft; thus, they are not preferred when
executing certain tasks such as transport and long-distance reconnaissance\cite{Saggiani2004Rotary}. This motivates to improve range and payload of multirotor UAVs\cite{Hassanalian2017Classifications}.

\begin{figure}[tbp]
	\includegraphics[
	width=12cm]{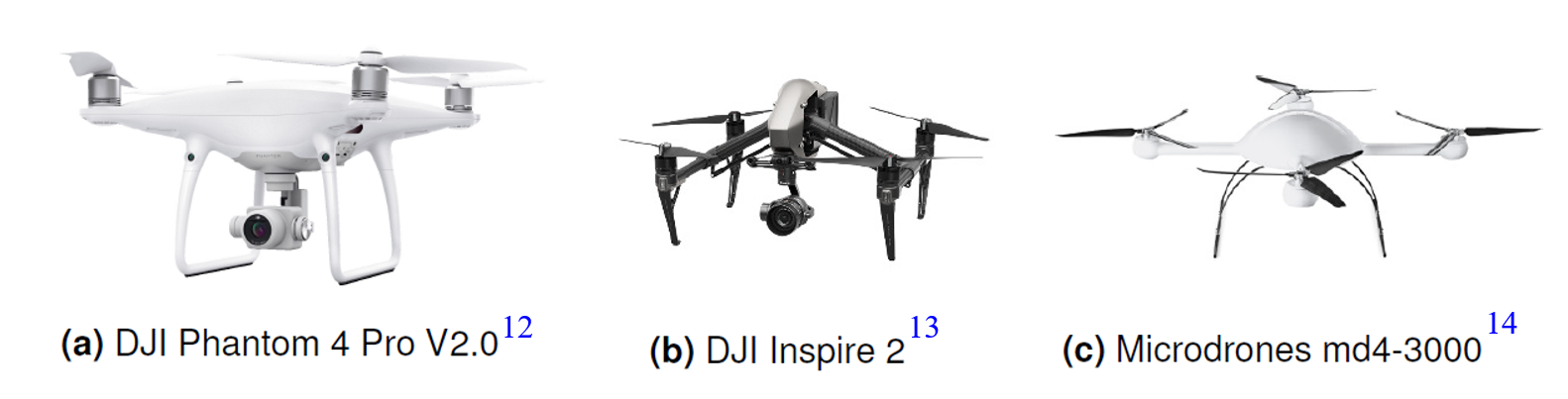}
\caption{Some multirotor UAV products with aerodynamic optimization}
\label{fig1:main}
\end{figure}

The general method to do this is to optimize propulsion
systems. Dai et al. \cite{8601389,shi2017practical,dai2018efficiency}
proposed an analytical design optimization method for electric propulsion
systems of multirotor UAVs. Magnussen et al. \cite{6935598} proposed a
design optimization method considering the number of actuators. Deters and
Selig \cite{deters2008static} and Ol et al. \cite{ol2008analytical}
contributed to characterize and optimize propeller performance. In addition
to propulsion system optimization, aerodynamic optimization of fuselage is
an effective way to improve range and payload. However, to the best of our
knowledge, there are limited academic works on aerodynamic optimization of
fuselage for multirotor UAVs. Hwang et al. \cite{doi:10.2514/1.C032828}
conducted a numerical study of aerodynamic performance of multirotor UAVs,
Bannwarth et al. \cite{doi:10.2514/1.J057165} built a novel multirotor UAV
aerodynamic model; however, they did not carry out the optimization
research. Compared with the academic world, industries pay more attention to
aerodynamic optimization. Fig. \ref{fig1:main} shows a few multirotor UAV
products \cite{phantom, inspire, md4-3000} with aerodynamic
optimization. It is evident that engineers focus on cutting down drag;
however, it is known that for an aircraft, there is not only drag, but also
lift.

\begin{figure}[tbp]
	\includegraphics[
	width=12cm]{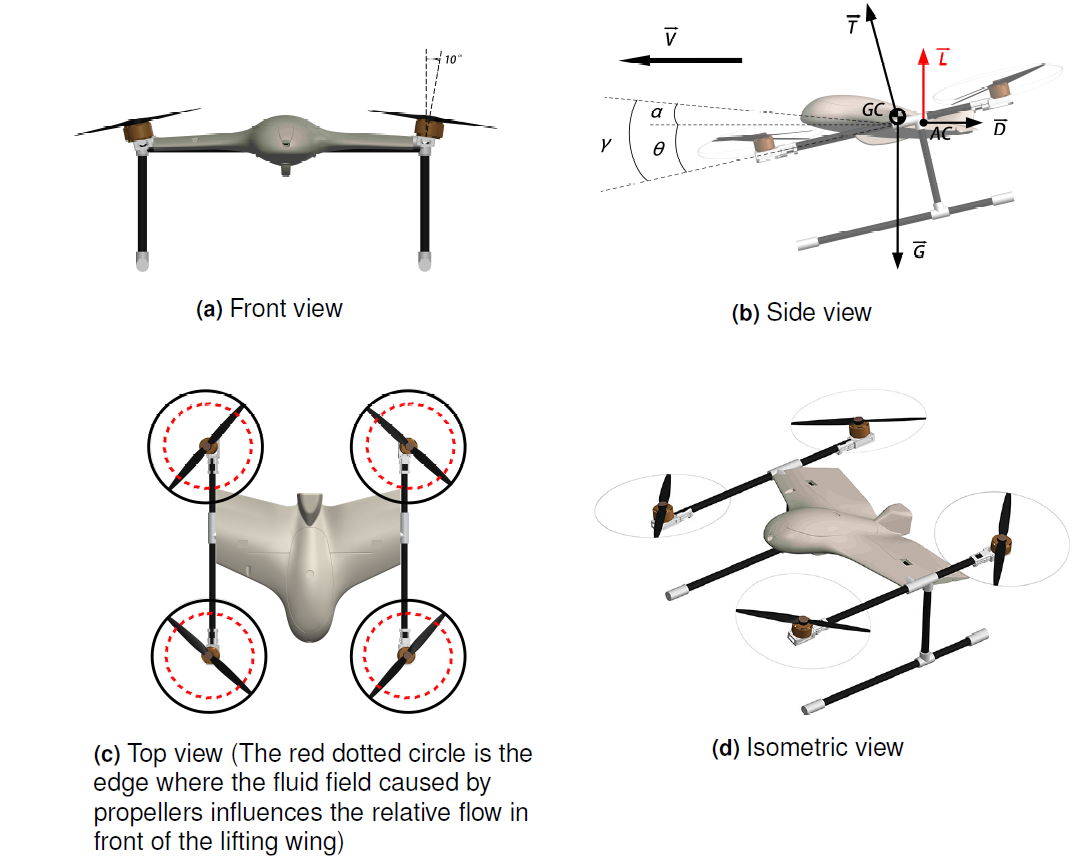}
\caption{3-View drawings of a lifting-wing multirotor UAV}
\label{fig3:main}
\end{figure}

As shown in Fig. \ref{fig3:main}, the key idea of our research is to study a
new type of multirotor UAVs, namely the \textit{lifting-wing multirotor UAVs},
which provides a multirotor UAV with a short wing installed at a specific
mounting angle. The lifting-wing multirotor UAV only has to tilt a specific
angle often smaller than 45 degrees to perform forward flight. After that,
both rotors and the lifting wing supply lift, thus reducing the energy
consumption and improving its range compared with the corresponding
multirotor UAV. Moreover, as shown in Fig. \ref{fig3:main}, it does not have a
tailfin. Instead, its function is replaced by the yaw control of the
multirotor UAV component. In order to increase the yaw control ability, the
axes of rotors do not point only upward any more (as shown in Fig. \ref
{fig3:main}(a)). This implies that the thrust component by rotors can change
the yaw directly rather than merely counting on the reaction torque of
rotors. From the above, the wind interference is significantly reduced on
the one hand; on the other hand, the yaw control ability is improved. As a
result, it can have better maneuverability and hover control to resist the
disturbance of wind than those by current hybrid UAVs. As a preliminary
study on the lifting-wing multirotor UAV, the design from the
aspects of aerodynamics, airframe configuration and wing's mounting angle will be discussed.
Also, the performance test is analyzed. Expectantly, the test results
show that the lifting wing saves 50.14\% power at the cruise speed (15 m/s).

The main contributions of this paper are: i) an analysis that aerodynamics of multiple propellers and the lifting wing are almost decoupled; ii)  a method to determine the mounted angle of the lifting wing; iii) the experimental study to show power saving.

\subsection{Comparison with other UAVs}

The lifting-wing multirotor UAV is a type of multirotor UAVs. But, it is necessary
to compare with existing fixed-wing Vertical/Short Take-Off and Landing
(V/STOL) UAVs, or hybrid UAVs in other words. V/STOL aerodynamic is
concerned primarily with the production of lift at low forward velocities
\cite{mccormick1999aerodynamics}. V/STOL UAVs in most time work as
fixed-wing UAVs. Thus, its hovering performance is considerably degraded by
the wind disturbance that is introduced by the wing \cite%
{doi:10.1177/1756829319869647}. According to a survey research \cite%
{SAEED201891}, hybrid UAVs with multiple rotors are classified into
multirotor tilt-rotor convertiplane, multirotor tilt-wing convertiplane,
multirotor dual-system convertiplane and multirotor tailsitter. Fig. \ref%
{fig2} shows these different kind of hybrid UAVs \cite%
{doi:10.2514/6.2014-0016, oner2012mathematical, aletky, google}. A comparison among different UAVs is listed as Table \ref{table1}.
As shown, our proposed design is a trade-off between the mutlicopter and the
fixed-wing airplane.
\begin{figure}[tbph]
\includegraphics[
		width=12cm]{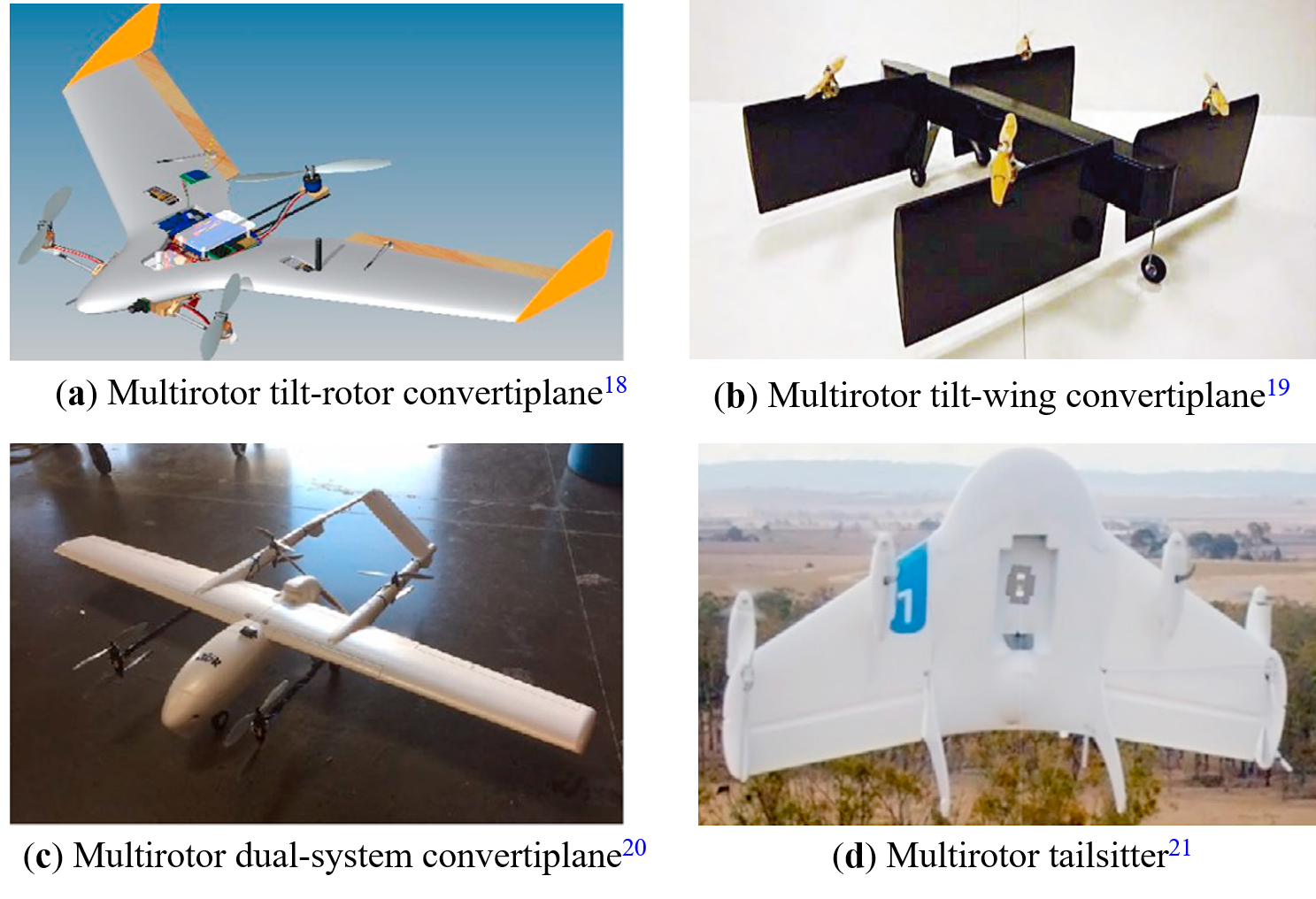}
\caption{Hybrid UAVs with multiple rotors}
\label{fig2}
\end{figure}
\begin{table}[tbph]
\caption{Comparison among different kinds of UAVs}
\includegraphics[
		width=12cm]{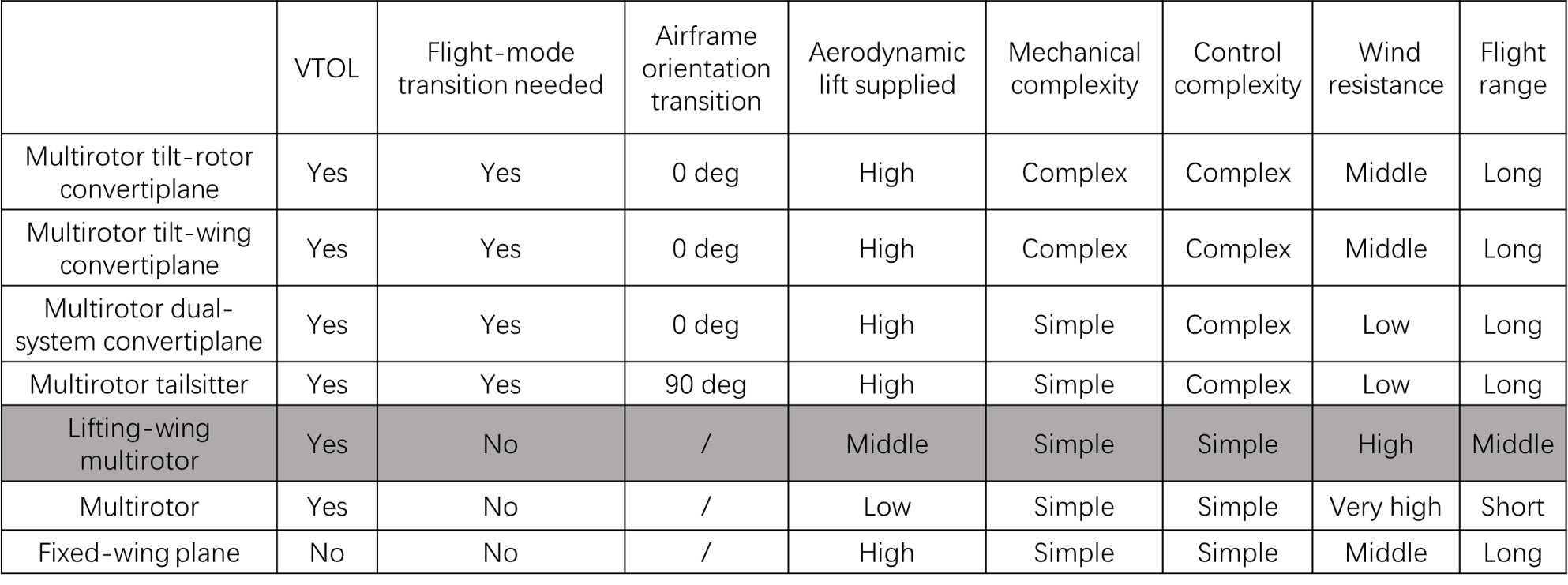}
\label{table1}
\end{table}

\section{Aerodynamics and Airframe Configurations}

\begin{figure}[tbph]
\includegraphics[
width=8.5cm]{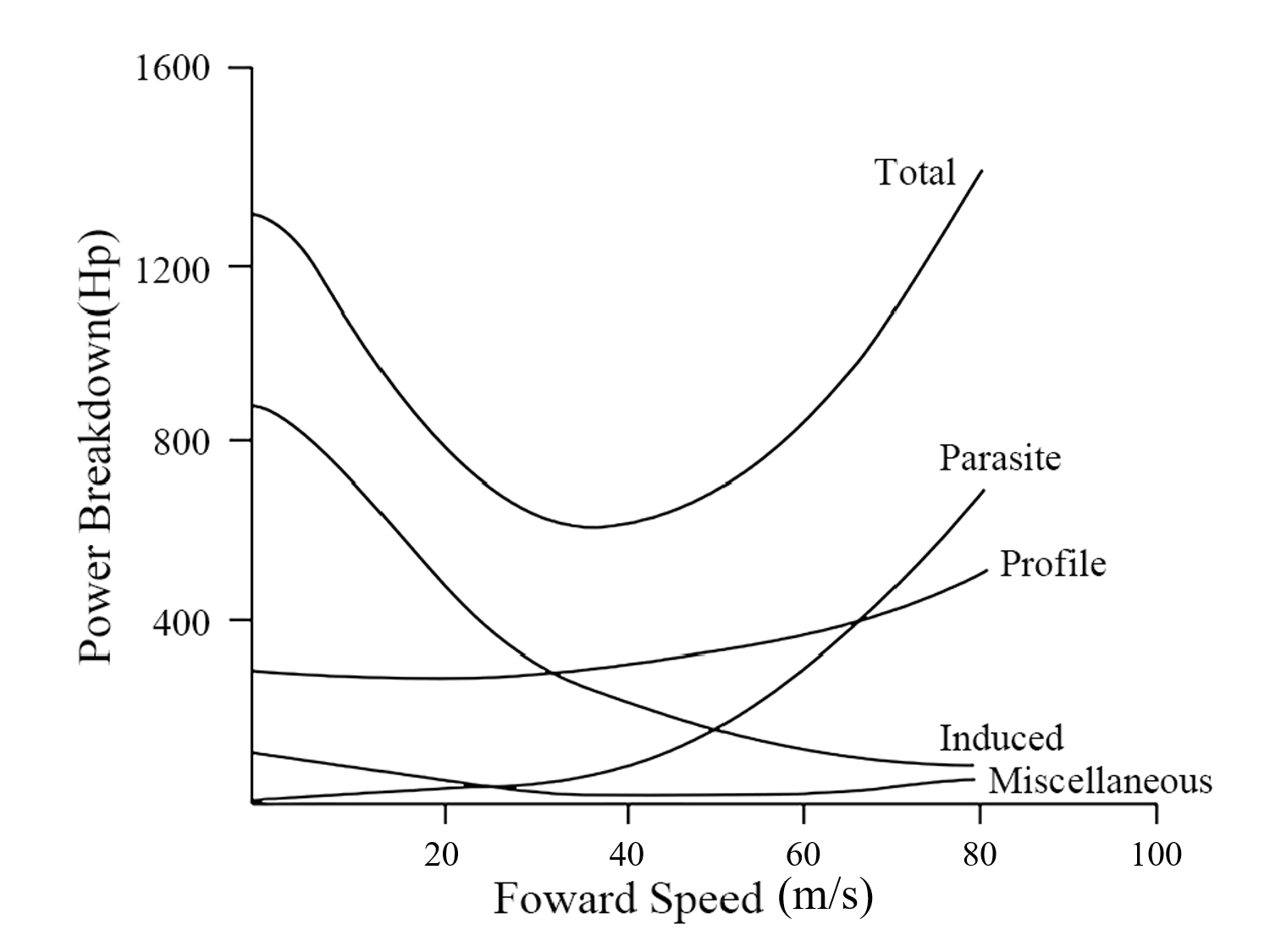}
\caption{Typical power breakdown for forward level flight of helicopter
\protect \cite{seddon2001basic}}
\label{fig4}
\end{figure}

In the introduction, it's shown that improving lift is more effective than cutting
down fuselage drag to improve range and payload. The opinion can be
explained through Fig. \ref{fig4}. The illustration, which comes from \cite%
{seddon2001basic}, shows that parasite (drag caused by fuselage) has very
little proportion under 20 m/s. Most of the power cost
comes from the propeller; thus, the effect of reducing fuselage drag is
limited.

Relatively, improving lift is an effective way, for it can reduce the need
of the component of propeller thrust in the vertical direction, which means
the component of the propeller thrust in the horizontal direction increases.
Therefore, the fuselage is designed as a lifting wing.

\begin{figure}[tbph]
\includegraphics[
width=8.5cm]{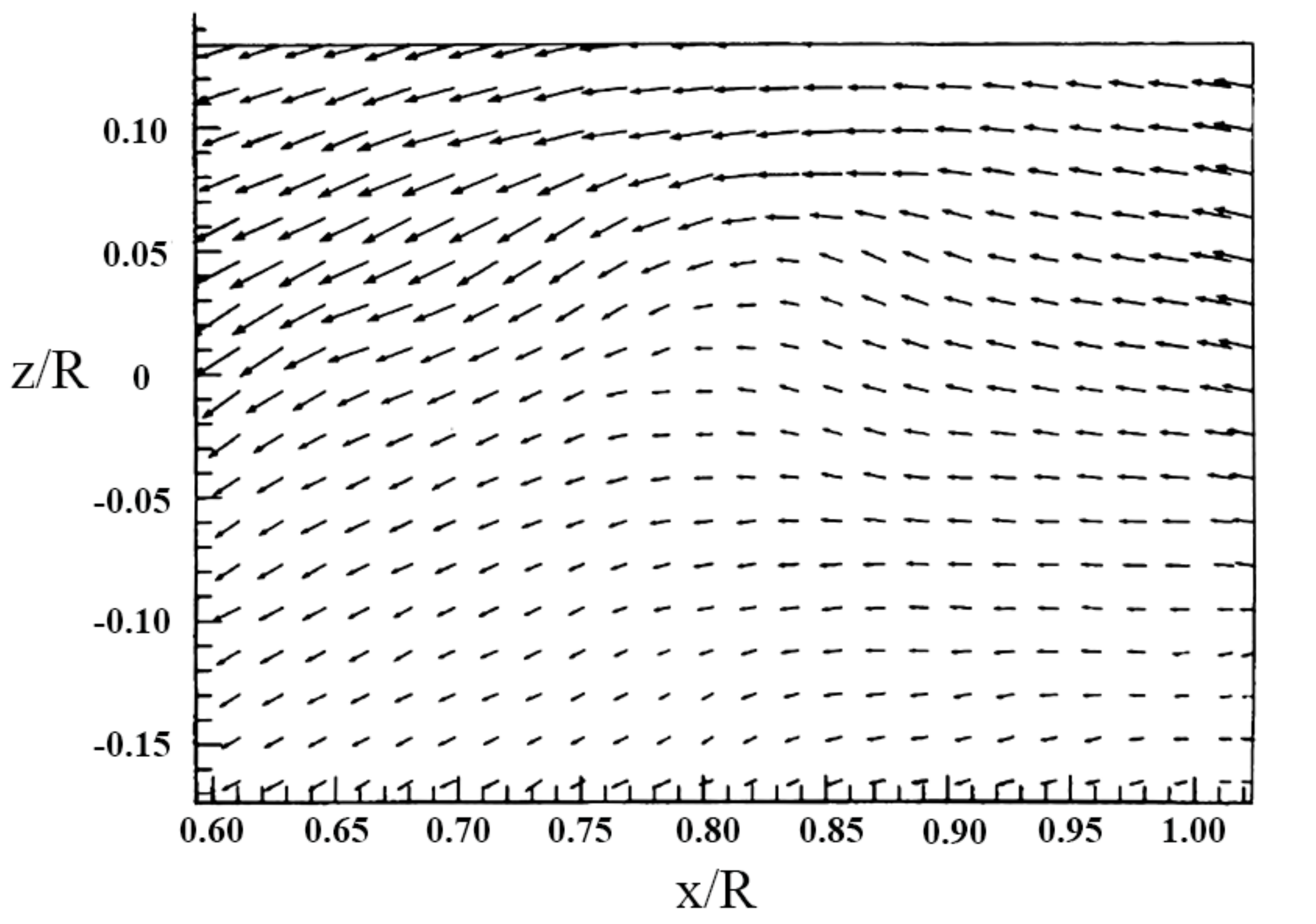}
\caption{Flow velocity field (advance ratio = 0.1), where x is the radial
distance toward the center of the propeller, z is the normal distance and R
is the radius of the propeller \protect \cite{jinghui2013heli}}
\label{fig5}
\end{figure}

An important question concerning the lifting wing design should be
addressed: Does the fluid field caused by propellers influence the relative
flow in front of the lifting wing? Fig. \ref{fig5} shows that the influence is little beyond 0.8 radius of propeller. And Fig. \ref{fig3:main}(c) shows that the position of the leading edge and the
trailing edge are both beyond 0.8 radius of propellers. Therefore, the wing
theory of fixed-wing aircraft can be used for the lifting wing, which makes
the design have rules to obey.

\begin{figure}[tbph]
\includegraphics[
width=10cm]{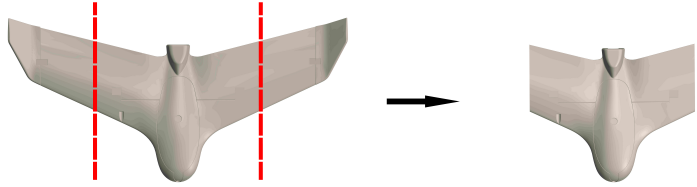}
\caption{The reshaping of Skywalker X5}
\label{fig6}
\end{figure}

For the experiment prototype, Skywalker X5 Blended Wing Body
aircraft is reshaped for the lifting wing. Fig. \ref{fig6} shows the manner in which the
wing is reshaped. The length of the wingspan is reduced and the winglets are
removed. Although in this way lift is reduced,additional force and moment
disturbances are reduced considerably, thus, achieving a trade-off between
range and wind resistance.

The yawing control moment of the multirotor UAV is caused by the air resistance moment of the propeller rotation. Therefore, the yawing control moment is weaker than the pitching and rolling control moments that are caused by different thrusts. Considering that the lifting wing will lead to an additional yawing moment when meeting with a crosswind, the yawing control moment should be improved. Therefore, the propellers in the prototype are tilted $10^{\circ}$  fixedly around two arms respectively, as shown in Fig. \ref{fig3:main}(a). Hence, the different thrusts lead to yawing control moment, improving the control performance.

\section{Mounting Angle Optimization}
The mounting angle is a term in fixed-wing aircraft, which is the angle
between the chord line of the wing and a reference axis along the fuselage
\cite{phillips2004mechanics}. For our proposed design, the mounting angle $%
\gamma$ also exists, which relates the two key angles, angle of attack $%
\alpha$ that decides the lift force, and pitch angle $\theta$ that decides
the ratio of the vertical components of thrust to the horizontal components.
Their relationship is shown as Equation (\ref{eq.1}) and Fig. \ref{fig3:main}(b).
\begin{equation}
\alpha=\gamma-\theta.  \label{eq.1}
\end{equation}

In this section, the mounting angle is optimized and the cruise
speed is determined for the purpose of obtaining the longest range.

\subsection{Optimization Model}

The optimization model is based on the assumption that there is no wind, and
the airframe is perfectly symmetric. The model considers the forward flight.
Thus, the roll moment, yaw moment, and lateral force can be neglected.
Therefore, the 3D dynamics can be simplified to 2D dynamics 
\begin{align}
&\sum \limits_{i=1}^{n}T_i\cos \theta+\frac{1}{2}{\rho}V^{2}SC_{L}(\alpha)=mg
\label{eq.2} \\
&\sum \limits_{i=1}^{n}T_i\sin \theta-\frac{1}{2}{\rho}V^{2}SC_{D}(\alpha)=0
\label{eq.3} \\
&M_{control}=M_{air}.  \label{eq.4}
\end{align}
where $T_i$ is the thrust magnitude for one propeller, $\rho$ is the air density, $V$ is the airspeed magnitude, $S$ is the reference area, $C_L$ is lift coefficient, $C_D$ drag coefficient, $m$ is the mass of the aircraft and g is the gravitational acceleration, $M_{control}$ is the control pitch moment and $M_{air}$ is the aerodynamic pitch moment.

Considering that four propellers can supply equal resultant force when
modifying the resultant moment, Equation (\ref{eq.4}) can be ignored in the
optimization problems because it is not an effective constraint.

Shastry et al. \cite{doi:10.2514/1.C034899} expressed the propeller thrust $%
T $ and torque $M_p$ in their simplified model as
\begin{align}
T&=\frac{C_T(N,V_p){\rho}N^2D_p^4}{16}  \label{eq.5} \\
M_p&=\frac{C_M(N,V_p){\rho}N^2D_p^5}{32}  \label{eq.6}
\end{align}
where $C_T$ is the propeller thrust coefficient, and $C_M$ is the propeller
torque coefficient. Both $C_T$ and $C_M$ depend on the rotation speed $N$ and
air speed perpendicular to the propeller disk $V_p$. Without considering the
environment wind, $V_p$ can be expressed as
\begin{equation}
V_p=V\sin \theta.  \label{eq.7}
\end{equation}
Therefore Equations (\ref{eq.5}) (\ref{eq.6}) can be written as Equations (\ref{eq.8}%
) (\ref{eq.9}) for the $i^{th}$ propeller. 
\begin{align}
&T_i=T_i(N_i,V,\theta) \label{eq.8} \\
&M_i=M_i(N_i,V,\theta) \label{eq.9}
\end{align}

In addition to the force and moment equations, electrical
equations are also part of the constraints of the optimization. 

\begin{align}
&I_i=I_i(M_i) \label{eq.10} \\
&Q=\sum\limits_{i=1}^{n}I_it \label{eq.11}
\end{align}
where $I_i$ is the current of one electronic speed controller, $Q$ is the battery power capability, and $t$ is the flight duration.

$C_L(\alpha)$, $C_D(\alpha)$ and Equations (\ref{eq.8}) (\ref{eq.9}) (\ref{eq.10}) (\ref{eq.11}) are fitted according to experiment data, which is presented in detail in Appendix.

The objective function is $R=Vt$, and according to the constraint equations, the optimization model can be expressed as	
\begin{equation}
\begin{split}
&\text{Maximize} \qquad R=V(\gamma,\alpha)t(\gamma,\alpha) \\
&\text{subject to} \qquad \text{Equations} (\ref{eq.1})(\ref{eq.2}) (\ref{eq.3}) (\ref{eq.8}) (\ref{eq.9}) (\ref{eq.10}) (\ref{eq.11})\\ 
&\text{and} \qquad \gamma\in[0,{\gamma}_{max}],\alpha\in[0,{\alpha}_{max}] \label{eq.12}.
\end{split}
\end{equation}

\subsection{Optimization Solution}

Equation (\ref{eq.12}) is a nonlinear programming problem. Considering limited
mechanical assembly accuracy, the mounting angle cannot be very precise;
therefore, we use the method of exhaustion. To avoid a stall and consider the
pitch angle limit, we set the enumeration range from 0 to $18^{\circ}$ (the
stall attack angle), and the installation angle ranges from $0^{\circ}$ to $%
50^{\circ}$ degree. Therefore, 900 steps are conducted in the solution.

\begin{figure}[]
	\begin{subfigure}{\linewidth}
		\includegraphics[width=12cm]{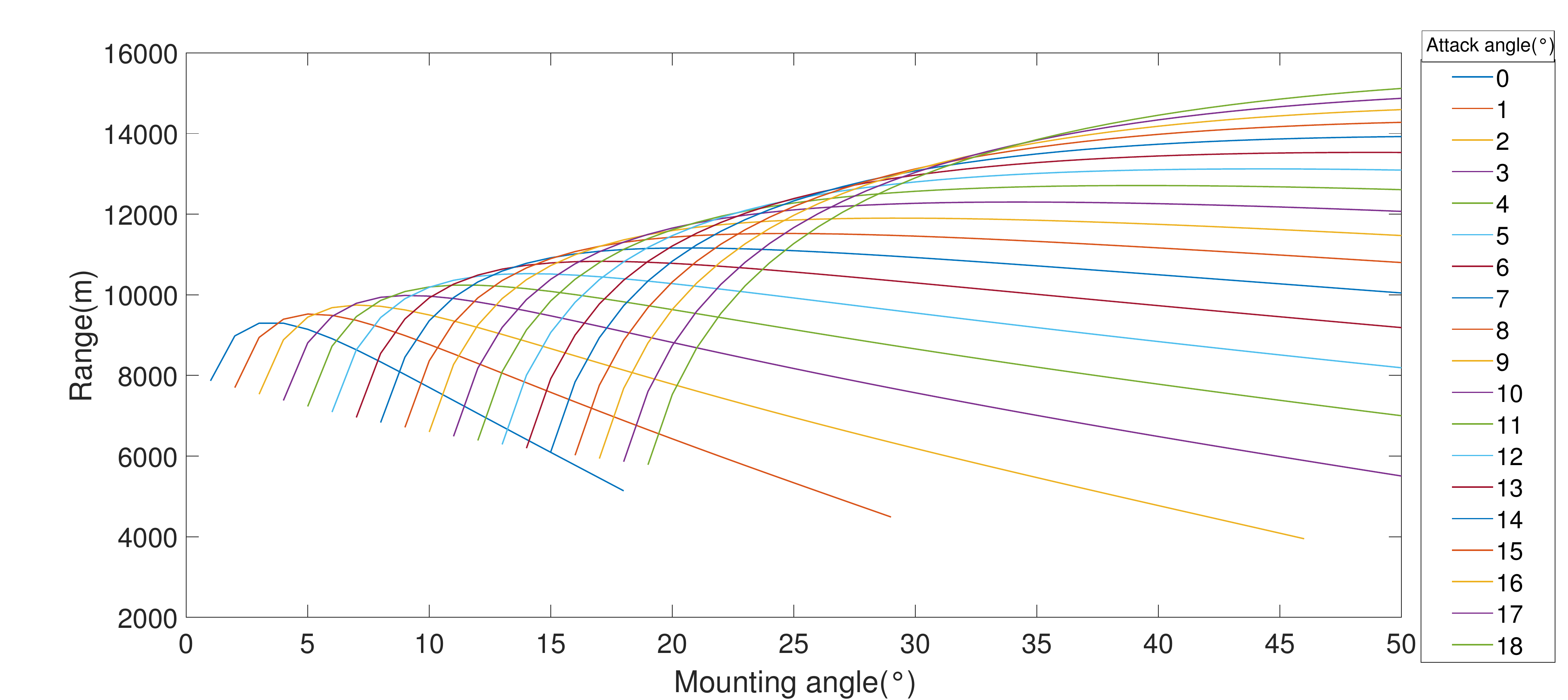}
		\caption{Original figure}
		\label{fig11:sub1}
	\end{subfigure}
	\begin{subfigure}{\linewidth}
		\includegraphics[width=12cm]{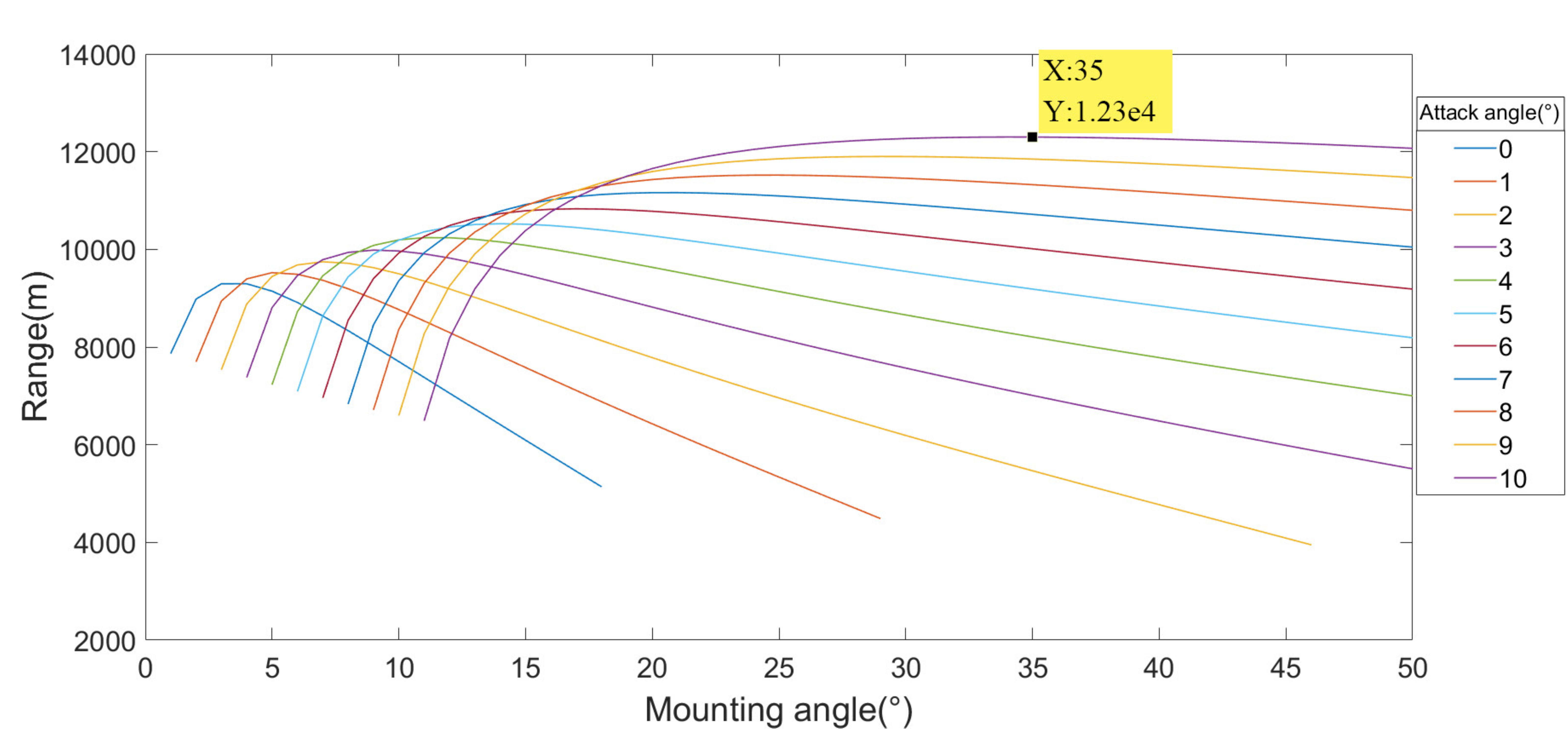}
		\caption{Figure with the limit of attack angle}
		\label{fig11:sub2}
	\end{subfigure}
	\caption{Range curve (varying with mounting angle and attack angle)}
	\label{fig11:main}
\end{figure}

Fig. \ref{fig11:main} shows the result. Fig. \ref{fig11:main}(a) is the
origin result, which shows that

\begin{enumerate}
	\item For a single curve, there is a maximum.
	
	\item As the attack angle increases, the maximum increases and the maximum
	point moves toward the right (therefore some maximum points are out of the
	x-axis range).
\end{enumerate}

To avoid a stall, we set $8^{\circ}$ as the safety margin of the attack angle.
Fig. \ref{fig11:main}(b) shows the limit of attack angle. The
flight range achieves its maximum (12.3
km) at $35^{\circ}$ mounting angle and $10^{\circ}$ attack angle. Under this
condition, the flight speed is 15.3 m/s. Therefore, we determine the mounting angle as $%
35^{\circ}$, and the cruise speed as 15 m/s.

\section{Experiment Verification}

In order to verify the proposed theory, a prototype was developed, and numerous outdoor flight experiments were conducted. A video which shows the experiments is available at

\url{https://youtu.be/YUjTbNmxSN4} 

or \url{http://rfly.buaa.edu.cn/index.html}.

\subsection{Experiment Settings}

\begin{figure}[htb]
\includegraphics[
width=8.5cm]{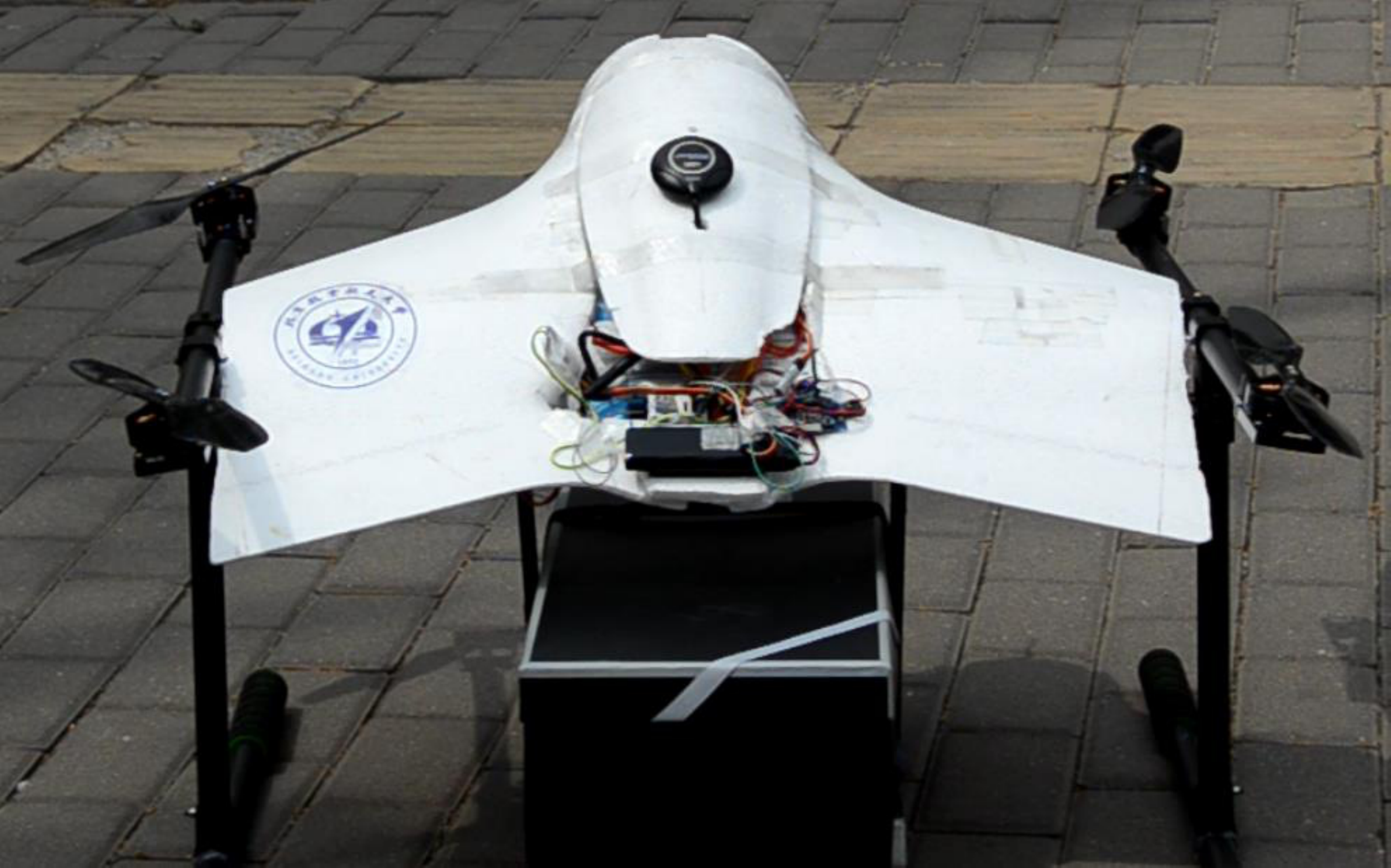}
\caption{Prototype carrying a package}
\label{fig7}
\end{figure}

\renewcommand{\thefootnote}{\fnsymbol{footnote}} Fig. \ref{fig7} shows the
prototype, whose weight is 2 kg, and diagonal size is 850 mm. The framework
is made of carbon fiber, and the lifting wing is mounted on the framework.
The flight controller is Pixhawk\footnote{http://pixhawk.org} (open source
hardware) along with Ardupilot\footnote{http://ardupilot.org} (open source
 software). We control the lifting-wing multirotor UAV under the multirotor UAV 
 control mode by taking the aerodynamic force and moment as disturbance.  

The flight environment is shown in Google Earth in Fig. \ref{fig8:main}. The
flight distance is approximately one kilometer, which is sufficiently long
for the aircraft to take adequate number of samples. For the purpose of
quantitative research, all the experiments were conducted in slightly windy
conditions (less than 2 m/s)\footnote{
We conducted a qualitative wind resistant experiment under the condition of
Scale 5 wind. The prototype succeeded in taking off, 10 m/s flight and
landing. The quantitative research of the wind resistant performance is our
future work.}. We analyze the flight performance including the control
performance and power consumption by analyzing the flight logs stored in the
controller.

\subsection{Control Performance Test}

\begin{figure}[htb]
\begin{subfigure}{\linewidth}
\includegraphics[width=8.5cm]{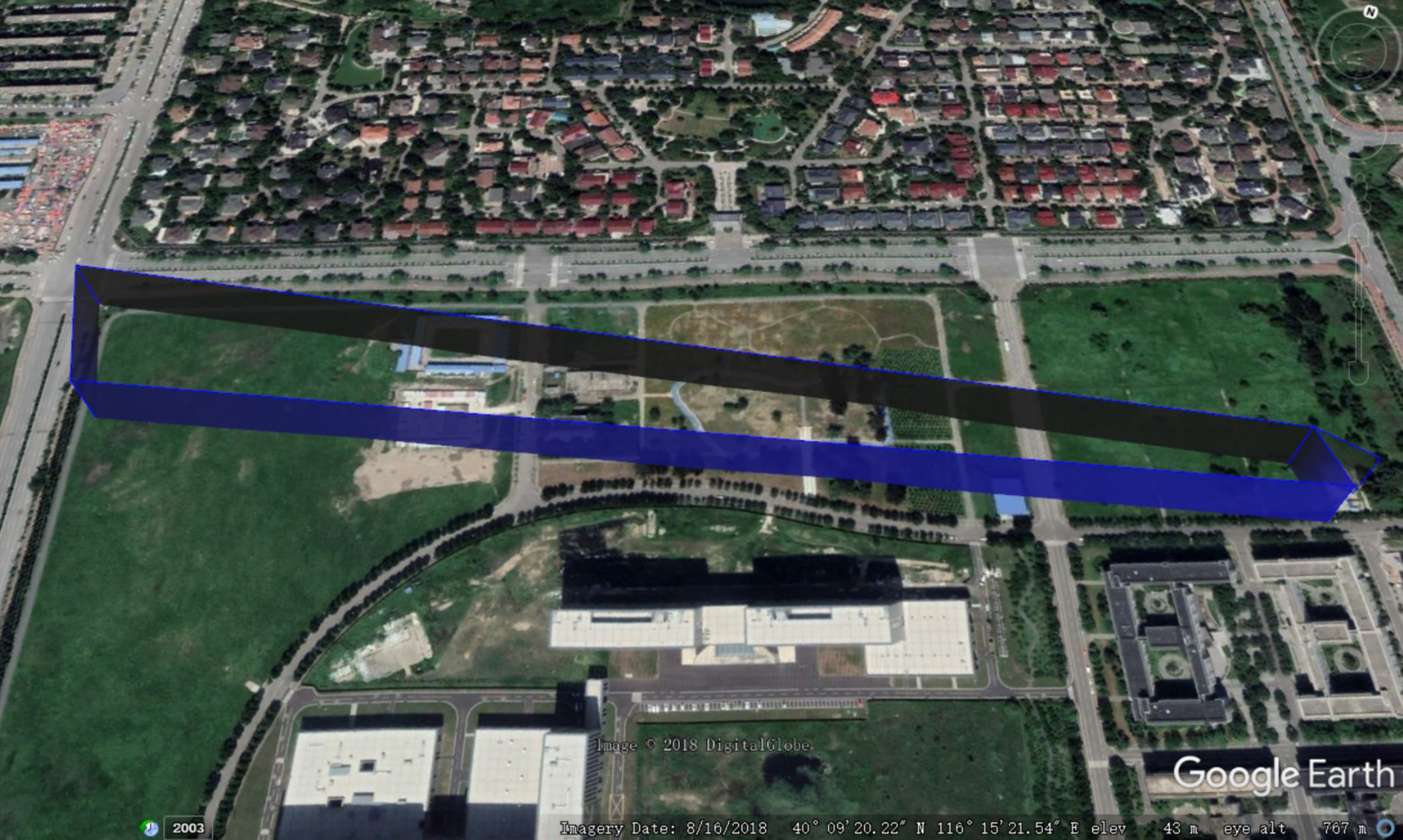}
\caption{Desired trajectory}
\label{fig8:sub1}
\end{subfigure}
\begin{subfigure}{\linewidth}
\includegraphics[width=8.5cm]{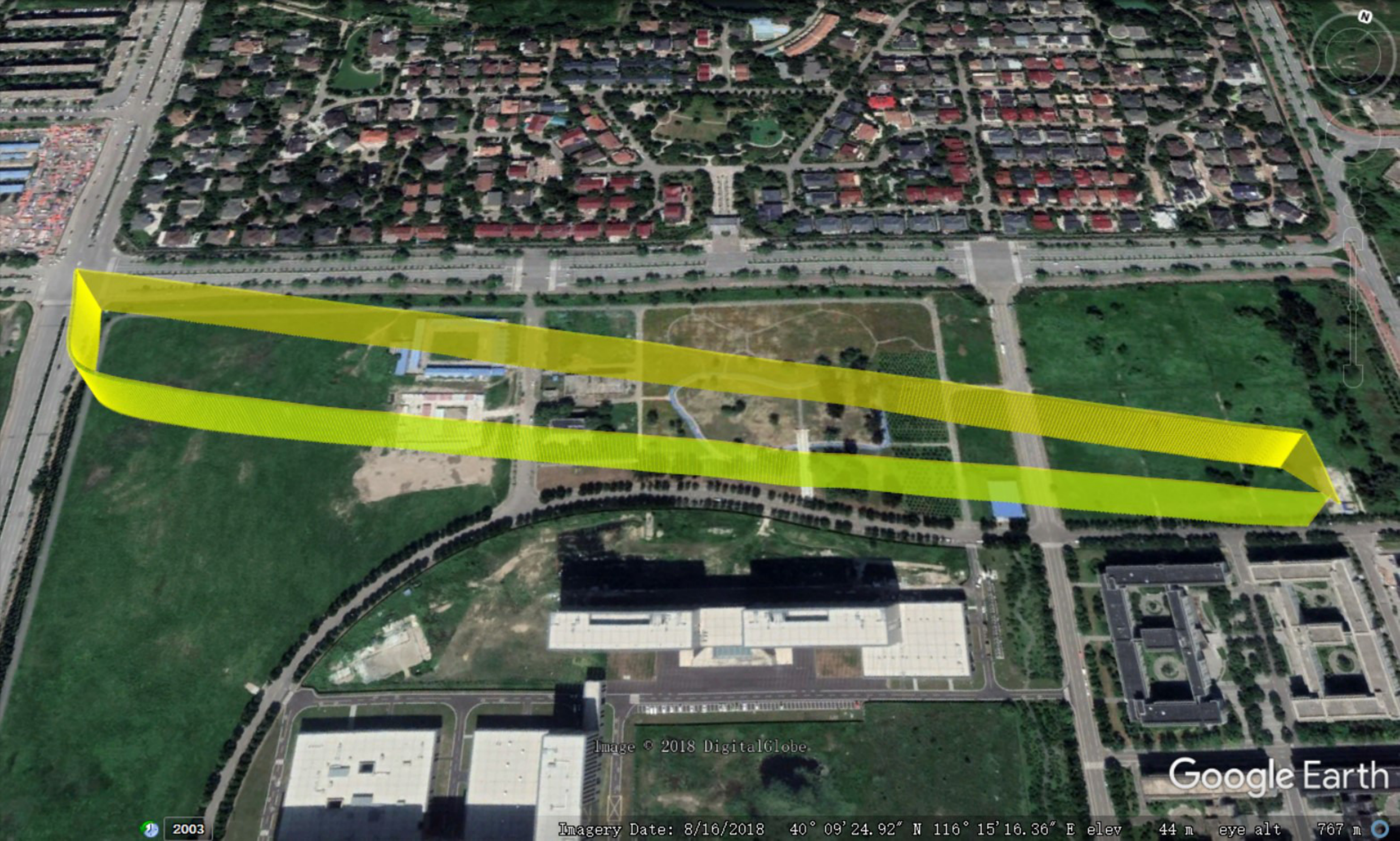}
\caption{Tracking trajectory}
\label{fig8:sub2}
\end{subfigure}
\caption{Trajectory tracking performance test}
\label{fig8:main}
\end{figure}

\begin{figure}[htb]
\includegraphics[
width=12.0cm]{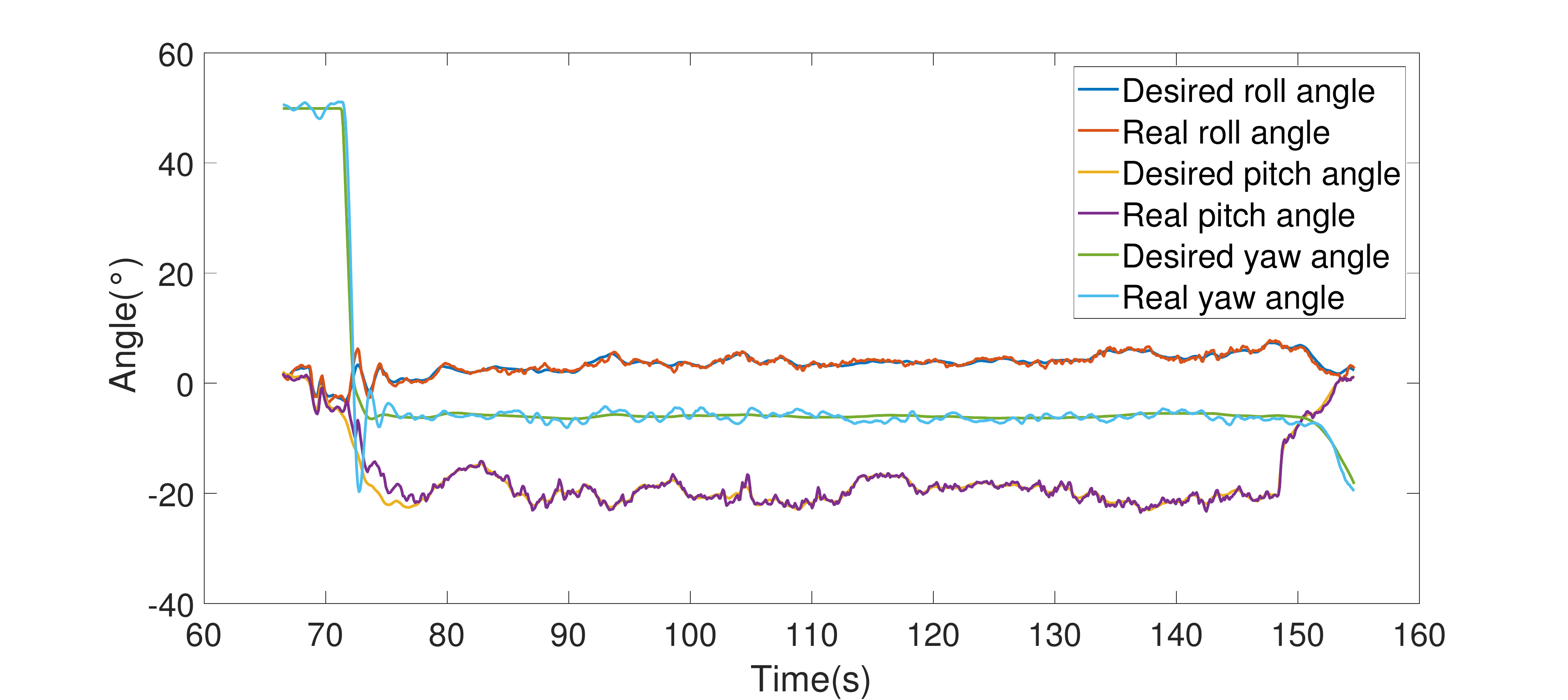}
\caption{Attitude tracking performance test}
\label{fig9}
\end{figure}

In the current flight mission, the flight speed is under 20 m/s, so the
additional aerodynamic force and moment can be considered as environment
disturbance. Therefore, the prototype is armed with the conventional
multirotor UAV controller which works well. Fig.\ref{fig8:main} shows that the
prototype tracks the desired trajectory well. Furthermore, Fig. \ref{fig9}
shows that the three attitude angles are tracked well during the 15 m/s
flight (including the adjustment period).

\subsection{Power Consumption Test}

\begin{figure}[htb]
\includegraphics[
width=8.5cm]{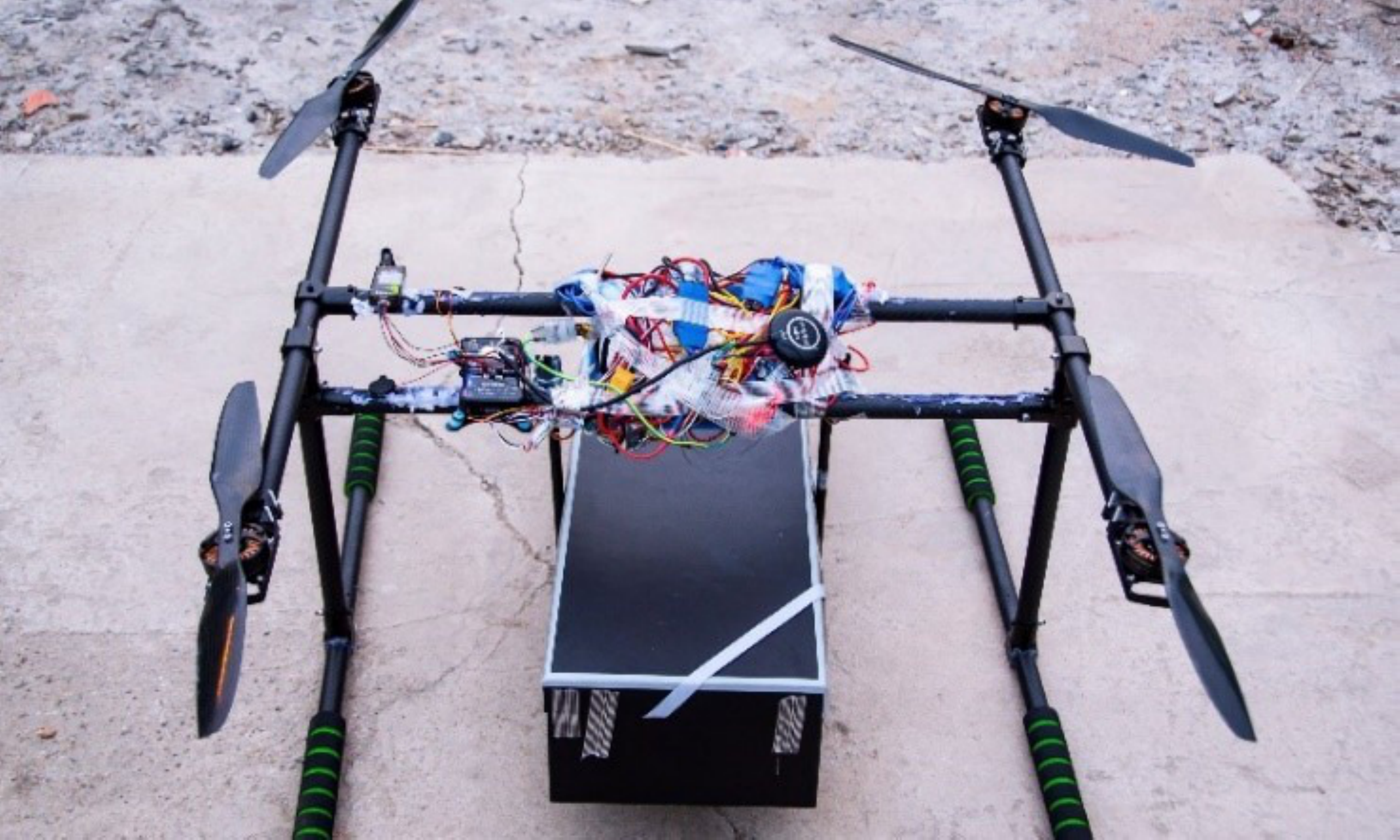}
\caption{Control Arm}
\label{fig10}
\end{figure}

To test the power consumption, which is the key to the performance of our
proposed design, we conducted a control experiment. The control arm, as Fig. 
\ref{fig10} shows, is a conventional multirotor UAV. For scientific control, it
is the same as the experiment arm (the prototype), except that it does not
have a lifting wing.

Table \ref{table2} compares the power consumption of the experiment arm with
that of the control arm. The real-time power is obtained from the flight
logs. The greater the flight speed is, the larger percent of power is saved
by the lifting wing. At 15 m/s (cruise speed), it saves 50.14\% power .

\begin{table}[htb]
\caption{Power consumption comparison}
\label{table2}
\begin{tabular}{|c|c|c|c|}
\hline
Flight speed & Power of control arm & Power of experiment arm & Power Save
\\ \hline
5 m/s & 2.436 mAh/s & 2.351 mAh/s & 3.49\% \\ \hline
10 m/s & 2.735 mAh/s & 1.921 mAh/s & 29.76\% \\ \hline
15 m/s(cruise speed) & 5.287 mAh/s & 2.636 mAh/s & 50.14\% \\ \hline
\end{tabular}%
\end{table}

\section{Conclusion}

The lifting wing design for multirotor UAVs is presented. The lifting wing provides additional lift force, which saves power, thus increasing the flight range. It is demonstrated that the aerodynamics of multiple propellers and the lifting wing are almost decoupled. Moreover, the mounting angle is optimized to obtain
the maximum flight range and determine the cruise speed. The experiment test shows that the lifting wing design saves power, and the greater the flight speed, the larger the percent of power is saved. For the cruise speed of 15 m/s, the prototype saves 50.14\% power. In the current work, a conventional multirotor UAV controller is applied to the lifting-wing multirotor UAV. This control scheme works well in the current flight mission; however, its performance worsens when the flight speed is greater than 20 m/s. The future work will focus on exploiting the aerodynamic force and moment to achieve a better control performance by adding control surfaces and designing a new controller.

\section{Appendix}

\subsection{Obtaining $C_L(\protect \alpha)$ and $C_D(\protect \alpha)$}

The aerodynamic coefficients are obtained from the wind tunnel experiments
from \cite{doi:10.1177/1756829318813633}, which introduces a VTOL UAV that
also uses Skywalker X5 as the aerodynamic configurations. After the linear
fitting, the expressions are obtained as
\begin{align}
C_L&=0.08\alpha-0.24, (-8^{\circ}\leq \alpha \leq18^{\circ})  \label{eq.13}
\\
C_D&=0.01587\alpha+0.14, (-8^{\circ}\leq \alpha \leq18^{\circ}).
\label{eq.14}
\end{align}

\subsection{Obtaining $T=T(N,V,\protect \theta)$ and $M=M(N,V,\protect%
\theta)$}

The propeller data is obtained from the APC Propeller official website%
\footnotemark[1]. The dataset contains different types of data, among which $%
V_p$, $N$, $T$, and $M$ are required. Equations (\ref{eq.8}) (\ref{eq.9}) are
expressed as Equations (\ref{eq.15}) (\ref{eq.16}). The coefficients of
correlation of the two fitting are 0.99993 and 0.99999. \footnotetext[1]{%
https://www.apcprop.com/files}
\begin{equation}
\begin{aligned}
T=&9.397\times10^{-2}+1.652\times10^{-3}-4.175\times10^{-5}N\\
&-7.915\times10^{-4}V_p^2-1.159\times10^{-5}V_p{N}+1.498\times10^{-7}N^2
\label{eq.15} \end{aligned}
\end{equation}
\begin{equation}
\begin{aligned}
M=&7.57\times10^{-2}+1.984\times10^{-2}V_p-2.466\times10^{-5}N-1.986%
\times10^{-3}N^2\\
&-5.308\times10^{-6}V_p\times{N}+1.275\times10^{-7}N^2-1.146\times10^{-5}N^3%
\\ &+1.562\times10^{-7}V_p^2\times{N}+1.227\times10^{-10}V_p\times{N}^2.
\label{eq.16} \end{aligned}
\end{equation}

\subsection{Obtaining $I=I(M)$}

The propulsion system experiment measurement was conducted using RCbenchmark
Series 1580 Thrust Stand and Dynamometer\footnotemark[1]. \footnotetext[1]{%
https://www.rcbenchmark.com/products/dynamometer-series-1580}

We conducted four samplings with different rotation speeds and fitted the
data with the quadratic function. The coefficient of the correlation is
0.9997, and the expression is as follows.
\begin{equation}
I=73.05M^2+12.15M-0.511  \label{eq.17}
\end{equation}

\bibliographystyle{SageV}
\bibliography{citation_mav}

\end{document}